# The silencing of neuronal activity by noise and the phenomenon of inverse stochastic resonance

B.S. Gutkin [1], J. Jost[2], H.C. Tuckwell*[2]

**Neurons in the central nervous system are affected by complex and noisy signals due to fluctuations in their cellular environment and in the inputs they receive from many other cells [1,2]. Such noise usually increases the probability that a neuron will send out a signal to its target cells [2-5]. In stochastic resonance, which occurs in many physical and biological systems, an optimal response is found at a particular noise amplitude [6-9]. We have found that in a classical neuronal model the opposite can occur - that noise can subdue or turn off repetitive neuronal activity in both single cells and networks of cells. Recent experiments on regularly firing neurons with noisy inputs confirm these predictions [10,11]. Surprisingly, we find that in some cases there is a noise level at which the response is a minimum, a phenomenon which is called *inverse stochastic resonance*. Suppression of rhythmic behavior by noise and inverse stochastic resonance are predicted to occur not only in neuronal systems but more generally in diverse nonlinear dynamical systems where a stable limit cycle is attainable from a stable rest state.**

In the central nervous system, neurons are embedded in complex networks of other neurons and glial cells [12,13]. They receive input signals from many other nerve cells through thousands of excitatory and inhibitory synapses at unpredictable or random times [14,15]. Many aspects of the transmission of signals in the nervous system are stochastic and the transmission process is itself unreliable or stochastic[16]. In order for a neuron to send out a signal, called an action potential, it must receive sufficient net excitation (over inhibition) in a small enough time interval. Technically this means that the current or voltage distribution in the cell must pass through some threshold condition [17]. Once the threshold is reached, self-exciting processes lead to the emission of an action potential.

___

1. Group for Neural Theory, Départément des Etudes Cognitives, Ecole Normale Supérieure, 3 rue d'Ulm, 75005 Paris, France.  2. Max Planck Institute for Mathematics in the Sciences, Inselstr. 22, 04103 Leipzig, Germany. *Corresponding author: tuckwell@mis.mpg.de

For many years the responses of neurons to input currents, whether injected or synaptic, have been investigated experimentally and theoretically[18]. Although it had long been realized that noise may accelerate neuronal activity [2-5], we have found that noise, especially of small amplitude, can decrease firing rates and even stop neuronal activity altogether, a finding which has been recently confirmed experimentally [10]. Furthermore, even though several recent investigations have demonstrated the phenomenon of stochastic resonance [6,7,9] in which a measure of response, such as a signal to noise ratio, rises to a maximum and then decreases as the noise amplitude is increased, we have found that the opposite behavior can occur; that is, as the noise increases from zero the response undergoes a minimum. Since the behavior is the reverse of that in stochastic resonance this new phenomenon can be called *inverse stochastic resonance*.

To study neuronal response to signals with noise we use the classical Hodgkin-Huxley model [19] which is capable of reproducing spiking properties which are similar to those of real neurons [17]. While cortical neurons present an intricate structure, with synapses distributed over an extensive dendritic tree (Fig1A left panel), for the purposes of our study we ignore the influence of such structure and consider a point neuron model. We also employ a simplified model of the signal and the noise: the signal is a constant time-independent current input to the cell while the noise is a diffusion approximation to randomly timed synaptic inputs [20].

Figure 1B shows the voltage responses of such a model neuron to input currents with various noise levels. The incoming signal has mean of strength µ and a noisy component of amplitude s. Without noise (top left record) there is, for the value of the steady input, µ=6.6, a repetitive stream of output spikes, there being 8 in the time period shown. Adding noise makes the output sequence irregular. Moreover small amounts of noise can actually stop the spiking. In the cases shown, a small noise of s =0.1 stops the firing of action potentials after 5 spikes; a somewhat larger noise level of s =0.5 has an even greater effect and stops the spiking after just one spike. When the noise level is turned up to s =1.5, more spikes are emitted, there being 6 in the trial shown.

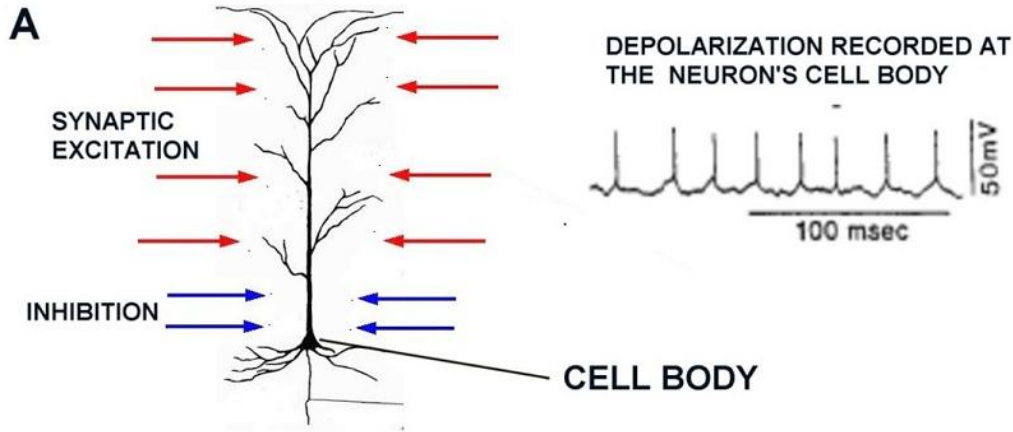

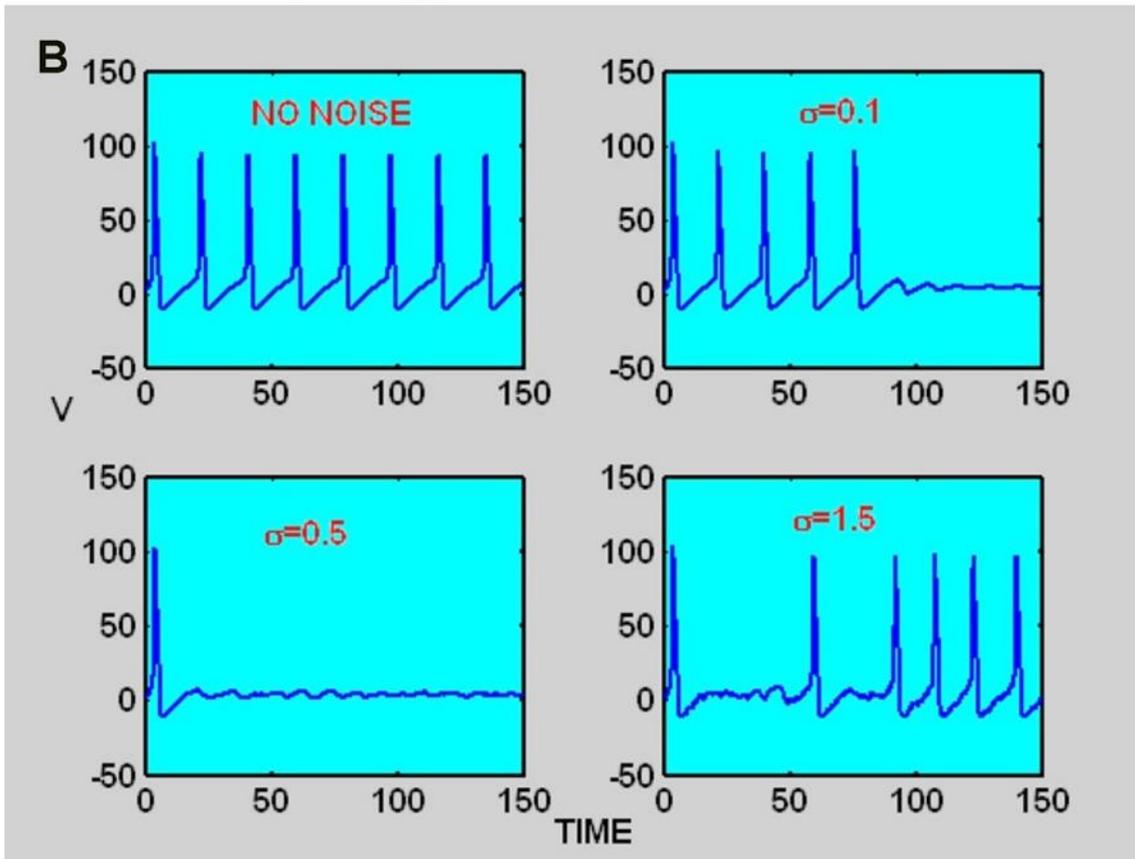

Figure 1.  A. At left is shown an outline of a pyramidal cell, the most frequently occurring type of neuron in the cerebral cortex. Arrowed lines represent schematically the arrival of excitation and inhibition, with blue for excitation and red for inhibition. On the right is shown a voltage recording showing 8 spikes emitted in about 200 msec. The spikes are seen to arise from a threshold.   B.  Plots of depolarization V  in  mV versus time (time unit .005 sec) showing spikes from a Hodgkin-Huxley model neuron with no noise and with noise of increasing magnitude.

In Figure 2 we show the results of a systematic exploration of the effects of noise on the regular spike train. The number of action potentials emitted by a Hodgkin-Huxley neuron (as typified by the records in Figure 1B) is plotted for input currents with various values of the mean current µ and of the noise level s. Without noise (s=0) there is a critical value of µ, $µ_c$ which is about 6.44, at which sustained repetitive firing occurs. For each data point with noise, 500 trials were performed.

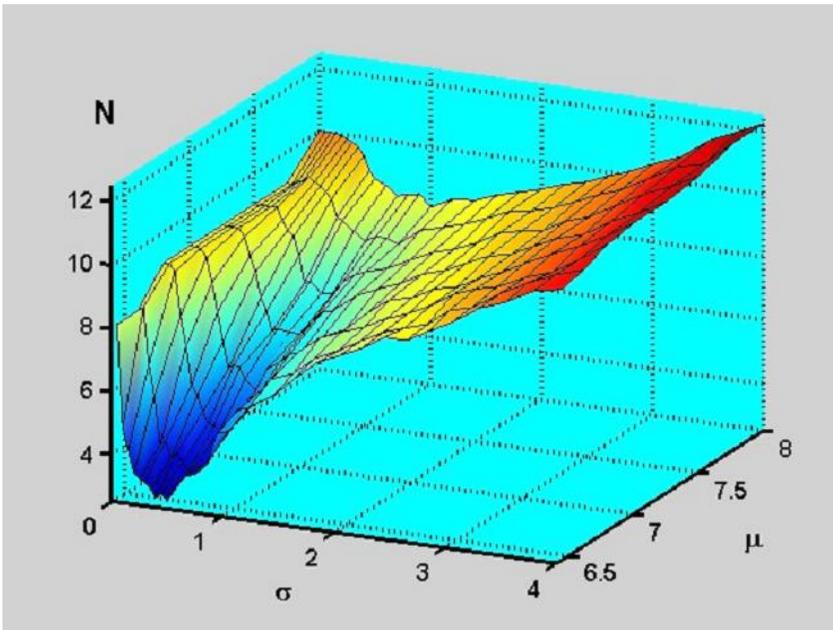

Figure 2. Inverse Stochastic Resonance in the Hodgkin-Huxley model. The mean number of spikes N versus mean µ and noise level s. The values of µ are above the critical value. Minima are clearly seen as s increases especially when µ is just greater than $µ_c$. Values of N for three particular values of µ are shown in the next Figure.

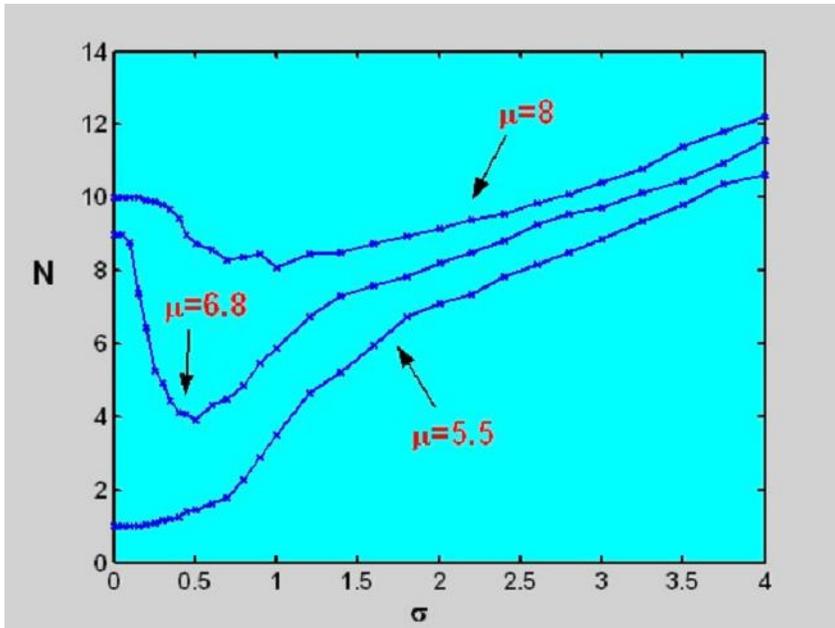

Figure 3.

Inverse stochastic resonance in the Hodgkin-Huxley model. Detailed plots of the results in Figure 2 for the three indicated values of µ showing number of spikes N versus noise level. Note the pronounced minimum when µ is 6.8, not far above the critical value for repetitive spiking.

Figure 3 shows how the presence of inverse stochastic resonance is dependent on the level of the deterministic input to the cell; that is, the mean incoming signal strength. To highlight the different behaviors for different values of µ, we have plotted N versus s at three values of µ. When µ=4 (not shown), there are no spikes without noise but increasing the noise level steadily increases N. As µ increases a small number of spikes occurs with no noise and the values of N again increase steadily as the noise level grows. However, as µ approaches the critical value, turning the noise on at first decreases and then increases the spiking activity. When µ is at or near the critical value there is a dramatic decline in spiking activity when the noise is switched on. This manifests as a pronounced minimum in the number of spikes as the noise increases, for values of µ just below $µ_c$ and just above $µ_c$. The minimum in the data for µ=6.8 is clearly discernible whereas there is no minimum at the smaller value µ=5.5 and a less noticeable minimum at the larger value µ=8.

The minimum in the response as the noise level increases through a certain value illustrates the phenomenon of inverse stochastic resonance. Note that the minimum in firing rate does not yield a minimum in the number of spikes divided by noise intensity, a quantity which could be called a signal to noise ratio. Our investigations of the silencing effects of noise on Hodgkin-Huxley neurons were motivated by our studies of pairs of coupled neurons of a different type[21] where a similar phenomenon was observed, namely that noise could cause the cessation of repetitive activity. We also investigated the effects of noise on repetitively firing pairs of coupled Hodgkin-Huxley neurons and obtained similar results and simulation of larger networks has yielded the same kind of behavior.

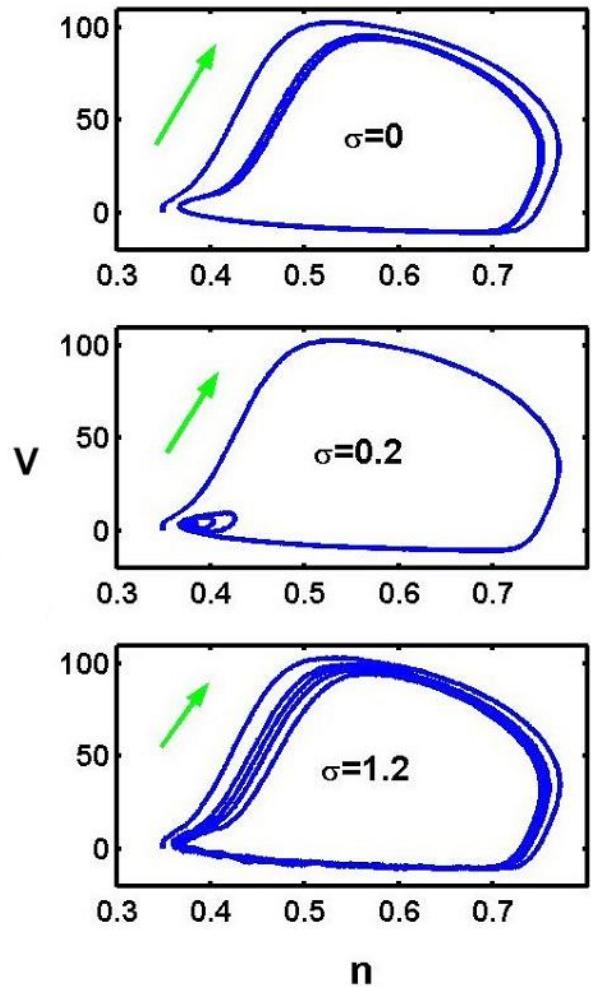

Figure 4. Voltage is plotted against potassium conductance variable n for results similar to those of Figure 1B. From top to bottom, no noise, s=0; middle, small noise, s=0.2 and bottom, large noise, s=1.2. The limit cycle is clearly seen in the noise-free case and the manner in which small noise, s=0.2, may switch the orbit away from the limit cycle.

The above results are explainable in terms of the behavior of the voltage and other variables on what are called stable limit cycles [22, 23] which occur when, for example, a neuron fires repetitively at the same frequency. Such a stable limit cycle in a dynamical system often appears by a bifurcation mechanism [22,23] when a parameter, like the input current strength µ in a Hodgkin-Huxley model, varies continuously and crosses some critical value. Just above that critical value, the basin of attraction of the limit cycle, that is, the region from which it is approached, is rather narrow. The stable limit cycle coexists then with one or more other attractors. In the Hodgkin-Huxley model, the only other attractor is a stable quiescent or resting state.  Noise can make the solutions leave the basin of attraction of the limit cycle for that of the quiescent state so that spiking ceases. When the noise is small, the solution will then typically stay quiescent for a very long time, but for larger noise there is then a considerable probability that the solutions get kicked back up to threshold so that spiking may resume. This may be followed by a period of relative silence and so on. This is illustrated in detail in Figure 4 where the voltage variable is plotted against the potassium conductance variable for µ=6.6, with no noise, s=0,  (top part), small noise, s=0.2 (middle) and large noise,  s=1.2 (bottom).

In the recent experiments on squid axon with noise [10], the effect of small noise has been likened to a switch. Thus, the functional significance of these effects of noise on rhythmic activity is that a very small disturbance can lead to a drastic change in behavior. In the brain, electrical activity is often broadly rhythmic, involving limit-cycles in both normal and epileptic activity [24, 25.] If such oscillations arise near a bifurcation point, then a small noisy signal could lead to the cessation of, or a sharp modification of, rhythmic activity. This is true also for impulsive disturbances, not necessarily ones composed of smooth noise as we have investigated.  Since stable limit cycles occur in dynamical systems in diverse fields, we expect to find that the phenomena of suppression of cyclic or rhythmic activity by noise and possibly nverse stochastic resonance will have widespread occurrence. For example, limit cycle activity is found in circadian rhythms [26], cardiology [27], cell kinetics and tumor growth [28,29] and oscillating neural networks [24,25] as well as in climatology, ecology and astrophysics. Although the phenomena we have described are of great interest, as indeed is stochastic resonance, their functional significance in neurobiological and other dynamical systems remains to be fully explored. Similar findings were reported in a heuristic nonlinear stochastic model of affective disorders [30]. It seems that these effects could sometimes arise as pathologies rather than normal conditions, as for example if cardiac pacemaker activity was affected adversely by noise.

______________________________________

**Acknowledgements**

BSG was supported by the CNRS and Marie Curie EXT "BIND".

**Author contributions**

The authors contributed approximately equally to this work which arose from a previous project on coupled neurons [21].